\documentclass[twocolumn,showpacs,prb,amsmath,amssymb,superscriptaddress]{revtex4}
\usepackage{graphicx}
\usepackage{array}

\begin{document}

\title{The topological magnetoelectric effect in semiconductor nanostructures: \\ quantum wells, wires, dots and rings}

\author{Josep Planelles}
\email{josep.planelles@uji.es}
\homepage{http://quimicaquantica.uji.es/}
\affiliation{Dept. de Qu\'imica F\'isica i Anal\'itica, Universitat Jaume I, 12080, Castell\'o, Spain}

\author{José L. Movilla}
\affiliation{Dept. d'Educaci\'o i Did\`actiques Espec\'ifiques, Universitat Jaume I, 12080, Castell\'o, Spain}

\author{Juan I. Climente}
\affiliation{Dept. de Qu\'imica F\'isica i Anal\'itica, Universitat Jaume I, 12080, Castell\'o, Spain}

\date{\today}

\begin{abstract}

Electrostatic charges placed near the interface between ordinary and topological insulators induce magnetic fields,
through the so-called topological magnetoelectric effect.
Here, we present a numerical implementation of the associated Maxwell equations.
The resulting model is simple, fast and quantitatively as accurate as the image charge method, 
but with the advantage of providing easy access to elaborate geometries when pursuing specific effects.
The model is used to study how magnetoelectric fields are influenced by the dimensions and the shape 
of the most common semiconductor nanostructures: quantum wells, quantum wires, quantum dots and quantum rings. 
Point-like charges give rise to magnetic fields of the order of mT, whose sign and spatial orientation is governed by the geometry of the nanostructure and the location of the charge. 
The results are rationalized in terms of the Hall currents induced on the surface, which constitute a simple yet valid framework for the deterministic design of magnetoelectric fields.

\end{abstract}

\maketitle

\section{Introduction}

Early works introducing axion electrodynamics in quantum field theory noticed that 
the ordinary Maxwell equations had to be complemented with additional terms, 
whereby static electric fields would induce magnetic polarization and, in turn,
static magnetic fields would induce charge polarization.\cite{WilczekPRL} 
This phenomenon, known as the magnetoelectric effect, has been later on predicted 
and observed in a number of condensed matter systems.\cite{NennoNR,YngAPX}
 One such system are topological insulators (TI).\cite{BernevigSCI,QiRMP}
Here, the magnetoelectric polarizability --responsible for the coupling of magnetic and electric fields-- 
is isotropic and takes quantized values in terms of the fine structure constant.\cite{MaciejkoPRL,VazifehPRB}
%
 This is in contrast to ordinary insulators (OI), where the polarizability is zero.
Experimental verifications of this effect have been recently reported in TI materials such as 
thin films of Cr$_x$(Bi$_{0.26}$Sb$_{0.74}$)$_{2-x}$Te$_3$\cite{OkadaNC} and
even-layer MnBi$_2$Te$_4,$\cite{LiuNM,GaoNAT} and potential technological applications 
are being proposed already.\cite{YueElec} 

An interesting manifestation of the topological magnetoelectric effect (TME) takes place in the presence 
of an interface between TI and OI. When an electric charge is placed
close to the interface, in addition to the usual dielectric polarization, a magnetoelectric 
polarization builds up on the TI surface. This triggers Hall currents on the interface,
which in turn generate local magnetic fields.\cite{QiRMP}

The TME has been well studied in the simplest heterostructure,
namely a planar interface between two semi-infinite bulks of TI and OI materials,
subject to the electric field generated by a point-like charge.
Because of the analogy with the dielectric mismatch effect, the theoretical models developed for 
such systems have often relied on the image charge method,\cite{Jackson_book,TakagaharaPRB93}
albeit extended for axionic Maxwell equations.\cite{QiRMP,YngAPX} 
The same method has been recently used to derive analytical solutions for the magnetoelectric fields
arising at the double interface of a TI slab embedded within an OI,\cite{YngAPX}
and the same structure but including all possible combinations of materials and external magnetizations.\cite{MoviXXX}
A few other geometries have been modeled too, including spheres with a centered charge\cite{FechnerPRB},
or semi-spherical cavities.\cite{CamposPLA} Equations for planar, spherical and cylindrical boundaries have been proposed
using the Green's matrix method\cite{MartinPRD}, but these become increasingly involved as the system symmetry is lowered.

Semiconductor nanostructures constitute a particularly rich playground to further investigate the properties
and prospects of the TME. When compared to micro and mesoscopic materials\cite{YngAPX}, 
the smaller dimensions make the induced magnetic fields potentially more relevant in affecting the properties of 
confined particles. Also, the fact that these particles are in the quantum regime, facilitates the experimental
observation of fine effects.\cite{FuhrerNAT,BayerPRB,ShornikovaNS}
In addition, the synthetic routes currently deployed in the fabrication of semiconductor
nanostructures enable the development of boundaries with a wide variety of controlled geometries.
Such structures can be placed in contact with TI of different materials and shapes.\cite{BernevigSCI,MalkovaPRB,KalesakiPRX,BeugelingNC,WanAM}
The curvature and symmetries of such interfaces can be expected to have a profound effect on the induced 
electromagnetic fields,\cite{OuelletPRD} which is worth exploring.

In this work, we report a systematic study on the effect of the nanostructure shape and size on
the magnetic field induced by electrical point charges. We focus on the most paradigmatic semiconductor 
nanostructures: quantum wells, quantum wires, quantum dots (spherical or cuboidal) and quantum rings,
embedded within TI media.
 Maxwell equations, includinc axionic terms, are solved by means of a simple and accessible numerical model
 based on finite elements method, which is shown to reproduce with great accuracy the results of image charge 
 methods for planar interfaces, but allows one to go beyond them and describe arbitrary shapes. 
 The results evidence that the curvature and position of the nanostructure boundaries, with respect to the
 source charge, can be engineered to either suppress or reinforce the induced magnetic field, and the field orientation 
 can be switched from roughly that of a magnetic dipole to complex, multipolar architectures.

\section{Theoretical Model}

Our starting point are the ordinary Mawxwell equations in Gaussian units:
\begin{eqnarray}
	\label{eq:M1}
	\boldsymbol{\nabla} \cdot \mathbf{D} &=& 4\pi \rho, \\
	\label{eq:M2}
	\boldsymbol{\nabla} \times \mathbf{E} &=& -\frac{1}{c}\,\frac{\partial \mathbf{B}}{\partial t}, \\
	\label{eq:M3}
	\boldsymbol{\nabla} \cdot \mathbf{B} &=& 0, \\
	\label{eq:M4}
	\boldsymbol{\nabla} \times \mathbf{H} &=& \frac{1}{c}\,\frac{\partial \mathbf{D}}{\partial t} + \frac{4\pi}{c}\,\mathbf{J}. 
\end{eqnarray}
\noindent The constitutive equations of displacement and magnetic fields are then modified 
to accommodate the axionic terms:\cite{QiRMP,MartinRuizIJMPA}
\begin{eqnarray}
	\label{eqD}
	\mathbf{D} &=& \epsilon\,\mathbf{E} - P\, \frac{\alpha \, \theta}{\pi}\, \mathbf{B}, \\
	\label{eqH}
	\mathbf{H} &=& \frac{\mathbf{B}}{\mu} + P\, \frac{\alpha \, \theta}{\pi} \mathbf{E}.
\end{eqnarray}
Here, $\epsilon$ and $\mu$ are the relative dielectric constant and magnetic permeability, respectively.
$\alpha=e^2/(\hbar c)$ is the fine structure constant, and $\theta$ the magnetoelectric polarizability.
In OI materials, $\theta=0$, so that Eqs.~(\ref{eqD},\ref{eqH}) become the usual expressions
of displacement, $\mathbf{D}=\epsilon \mathbf{E}$, and magnetic field $\mathbf{H}=\mathbf{B}/\mu$.
In bulk TI, however, $\theta=\pi$.\cite{MaciejkoPRL,VazifehPRB}
The sign is set by $P=sign[\mathbf{M} \cdot \mathbf{n}]$, which relates to the sense of rotation of the Hall conductance.
Here, $\mathbf{M}$ is an external magnetization lifting the time reversal symmetry near the interface,
and $\mathbf{n}$ the interface surface vector.\cite{QiRMP,MoviXXX} \\

In electrostatic systems, with no time-dependent field and absence of electrical currents $\mathbf{J}$, 
Eqs.~(\ref{eq:M2}) and (\ref{eq:M4}) become $\boldsymbol{\nabla} \times \mathbf{E} = 0$ and 
$\boldsymbol{\nabla} \times \mathbf{H} = 0$.
 This allows us to obtain the vector fields from the gradient of their scalar potentials,
$\mathbf{E}=-\boldsymbol{\nabla} V$ and $\mathbf{H} = -\boldsymbol{\nabla} W$. 
Then, Gauss law, Eq.~(\ref{eq:M1}), can be written as:
\begin{equation}
\label{eq:G1}
\boldsymbol{\nabla} \left[ -\left(\epsilon + \mu \, S^2 \right) \, \boldsymbol{\nabla} V
	+ P \, \mu \, S \boldsymbol{\nabla} W \right] = 4 \pi \rho
\end{equation}
\noindent with $S=\alpha \, \theta/\pi$.
In turn, Gauss law for magnetism, Eq.~(\ref{eq:M3}), can be written as:
\begin{equation}
\label{eq:G2}
\boldsymbol{\nabla} \left[ -\mu \boldsymbol{\nabla} W
	+ P \, \mu \, S \boldsymbol{\nabla} V \right] = 0.
\end{equation}

Equations (\ref{eq:G1}) and (\ref{eq:G2}) can be cast in matrix form as:
\begin{equation}
	\label{eq:S}
-\boldsymbol{\nabla}
\begin{pmatrix}
	\epsilon + \mu \, S^2 & -P \, \mu \, S \\
	-P \, \mu \, S & \mu \\
\end{pmatrix}
\,
\boldsymbol{\nabla}
\,
	\begin{pmatrix}
		V \\
		W 
	\end{pmatrix}
	=
	\begin{pmatrix}
		4 \, \pi \, \rho \\
		0
	\end{pmatrix}
\end{equation}\\
\noindent One can notice that off-diagonal terms vanish unless $\boldsymbol{\nabla} (\mu\,S) \neq 0$. 
That is, magnetoelectric coupling arises at the interface between materials with different 
magnetoelectric polarizability $\theta$.

The latter system of equations can be integrated for arbitrary 3D geometries using
standard finite element methods, by defining appropriate values of $\epsilon$, $\mu$ and $\theta$
in each region of the space.  We do so for a point-like electric charge of positive sign 
 ($Q=e$, with $e$ the fundamental unit of charge), 
 located at a position $\mathbf{r}_Q$, near the TI -- OI interface. 
 Boundary conditions are set at the edges of a large supercell surrounding the nanostructure,
far enough for the magnetoelectric potentials to be negligible, $V=0$ and $W=0$.
We shall see in the next section that this approximation, which greatly simplifies the model, 
enables accurate estimates of the fields.
 In our calculations, the integration of Eq.~(\ref{eq:S}) is carried out using the finite element routines of Comsol Multiphysics 4.2.
 The computing times range from tens of seconds to a few minutes on an ordinary PC.

 Having the electric and magnetic potentials, $V$ and $W$, we can obtain the magnetic field $\mathbf{B}$ as 
\begin{equation}
\mathbf{B}=-\boldsymbol{\nabla} U,
\end{equation}
\noindent where the associated potential is $ U = \mu \left( W - P S V  \right)$.\\

For a qualitative analysis, the magnetic field originating from a static source charge $\rho$ can 
be understood by replacing Eq.~(\ref{eqH}) into Ampere's circuital law, Eq.~(\ref{eq:M4}).
In the absence of time-dependent displacement $\mathbf{D}$, the equation can be written as:
\begin{equation}
\label{eq:Hall}
\boldsymbol{\nabla} \times \mathbf{H}' = \frac{4\pi}{c}\,\left(\mathbf{J}+\mathbf{J}_\theta \right). 
\end{equation}
\noindent with $\mathbf{H}'=\mathbf{B}/\mu$ and 
$\mathbf{J}_\theta=-\alpha \, c \,P\, \left( \boldsymbol{\nabla} \theta \times \mathbf{E} \right) / (4 \, \pi^2)$.
 It follows from Eq.~(\ref{eq:Hall}) that the electric field $\mathbf{E}$ generated by a source charge $\rho$ 
induces (Hall) currents $\mathbf{J}_\theta$ on the interface between materials with different $\theta$,
where $\boldsymbol{\nabla} \theta \neq 0$. These Hall currents, in turn, give rise to 
magnetic fields, $\mathbf{H}'$ and $\mathbf{B}$.

\section{Results}

We study the magnetic fields $\mathbf{B}$ resulting from a point charge $Q$, placed
near TI -- OI (semiconductor) interfaces with different geometries.
 Even though our numerical method is of general validity, we consider the specific case where the 
 nanostructure is made of an ordinary semiconductor material, surrounded by a TI.
 The opposite case, TI nanostructures embedded in OI, can be also found in experiments,\cite{BernevigSCI,DabardCM,GeiregatNM}
 but spatial confinement is expected to reduce the magnetoelectric polarizability $\theta$,\cite{PournaghaviPRB}
 and eventually suppress the topological behavior.\cite{BernevigSCI,DabardCM}

 Typical material parameters are used to describe ordinary semiconductors ($\mu_1=1$, $\epsilon_1=5$, $\theta_1=0$) 
 and TI, such as bulk Hg chalcogenides ($\mu_2=1$, $\epsilon_2=10$, $\theta_2=\pi$). 
 We assume the external magnetization $\mathbf{M}$ is such that $P=-1$ for all interfaces,
 which allows comparing with earlier works studying simple heterostructures.\cite{YngAPX}

\subsection{Planar interface}
\label{s:plane}

 \onecolumngrid

\begin{figure*}[t]
\includegraphics[width=12.0cm]{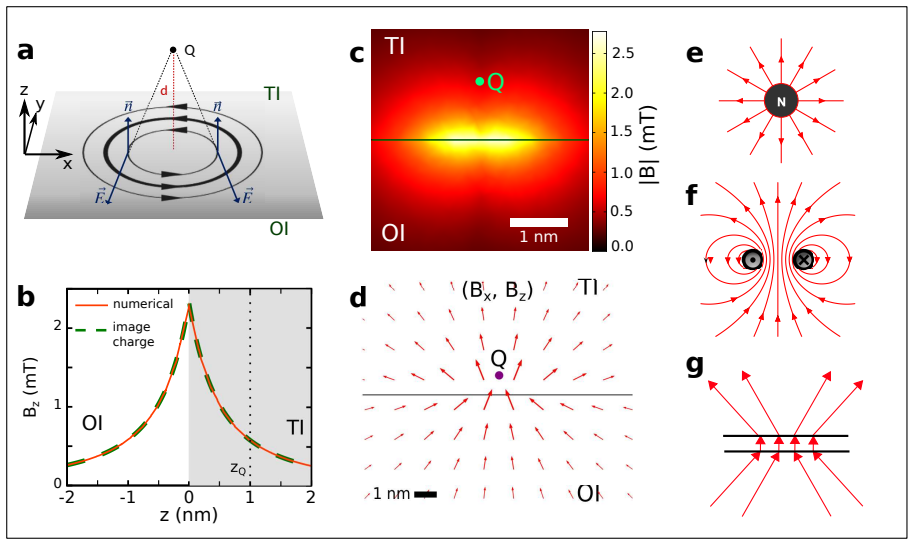}
\caption{TME on a planar interface.
(a) Schematic of the structure. The thickness of the circles represents
qualitatively the intensity of the Hall currents induced by the charge $Q$.
(b) $B_z$ projection of the magnetic field along the $z$ axis, calculated with 
numerical (solid line) and image charge (dashed line) methods.
(c) Absolute value of $\mathbf{B}$ on the $xz$ plane. The circle shows the position of the charge impurity.
(d) Logarithmic plot of the $(B_x,B_z)$ on the $xz$ plane. 
(e-g) Magnetic field distributions arising from a magnetic monopole (e), 
a circular current --quasi-dipole-- (f), and a Pearl vortex (g).
}
\label{fig1}
\end{figure*}

\twocolumngrid

 We start by considering the case where a semi-infinite bulk of OI (semiconductor) is placed on top
 of a semi-infinite bulk of TI, separated by a planar interface.
 The point charge $Q$ is located in the TI, at a distance $d$ from the interface, see Figure \ref{fig1}(a).
 We take $d=1$ nm, which is a plausible distance for controlled doping in nanostructured semiconductors.\cite{KoenraadNM,CapitaniNL} 
 
 Figure \ref{fig1}(b) shows the $B_z$ projection of the magnetic field induced along the $z$ axis, 
 which is perpendicular to the planar interface and crosses the charge position. 
 For this simple geometry, the field can be calculated analytically using the image charge method.\cite{QiRMP,YngAPX,MoviXXX}
 The comparison between numerical (solid line) and analytical (dashed line) calculations shows
 an excellent agreement, which illustrates the quantitative accuracy of the numerical method.
 The figure also reveals that $B_z$ has a maximum of few mT on the interface.
 Even if weak, the magnitude of $B$ is orders-of-magnitude greater than in millimetric structures,
 where values of fT are expected,\cite{FechnerPRB}
  and experimental detection of electronic effects under fields of similar magnitude have been reported in quantum nanostructures.\cite{FuhrerNAT}
 Yet, $B_z$ decays rapidly with the distance from the interface. 
 Along the $z$ axis, $B_z \propto (|z| + d)^{-2}$, with $z=0$ corresponding to the interface plane. 
 Remarkably, this is a magnetic monopole-like decay.\cite{QiRMP}
 The fast decay of the field can also be observed in Figure \ref{fig1}(c), which shows the absolute value $|\mathbf{B}|$ 
 on the $xz$ plane containing $Q$. The magnetic field is largely a local effect, arising in the vicinity of the interface under $Q$. 
 It follows from the fast decay of $B$ that the use of multiple charge dopants may extend the field domain, 
 but not its peak intensity, unless the concentration is very high ($>1$ nm$^{-3}$).
 
 Figure \ref{fig1}(d) is a logarithmic plot showing the magnetic field orientation for the same plane as Fig.~\ref{fig1}(c).
 It is clear from the figure that the orientation of $\mathbf{B}$ reflects neither a magnetic monopole (Fig.~\ref{fig1}(e)), 
 nor the quasi-dipole associated to a circular current (Fig.~\ref{fig1}(f)).
 It has been recently pointed out that, for sufficiently small $d$, the distribution in this setup 
 corresponds to a Pearl vortex instead (Fig.~\ref{fig1}(g)).\cite{NogueiraPRR}

 The deviation of the field orientation from that of a magnetic dipole can be understood by analyzing the Hall current, 
 $\mathbf{J}_\theta$.  As mentioned before, an electrical charge $\rho$ induces a current 
 $\mathbf{J}_\theta=\alpha \, c \, \left( \boldsymbol{\nabla} \theta \times \mathbf{E} \right) / (4 \, \pi^2)$.
 Because $\boldsymbol{\nabla} \theta = \pi\,\delta(r_i)\,\mathbf{n}$, with $r_i$ the interface coordinate and
 $\mathbf{n}$ a surface vector pointing towards the TI, 
 the sign and intensity of $\mathbf{J}_\theta$ are mainly given by the cross product $\mathbf{n} \times \mathbf{E}$.
 As can be seen in Figure \ref{fig1}(a), a point charge generates circular currents $\mathbf{J}_\theta$. 
 Each of the (infinite) loops behaves as a quasi-dipole, but their superposition does not.
 It is worth noting that the intensity of the currents is set by two factors: 
  
 (i) the distance from $Q$; since $|\mathbf{E}|$ scales as $1/|r-r_Q|^2$, currents will become weaker at longer distances.

 (ii) the angle between $\mathbf{n}$ and $\mathbf{E}$; the cross product becomes null when the two vectors are aligned
 (right below $Q$), and increase as they become orthogonal (longer distances).

 The trade-off between the previous effects leads to the most intense Hall currents taking place for radii $R = d$,
 where $\mathbf{n}$ and $\mathbf{E}$ form an angle of $\pi/4$. Weaker currents are expected closer or farther from $Q$, 
 as sketched by the different thickness of the lines in Figure \ref{fig1}(a). The dominating character of
 such loops impose a magnetic field distribution loosely resembling that of a dipole, albeit with clear deviations
 originating from the interferences with other loops. The absence of vortices around the inbound and outbound current 
 in Fig.~\ref{fig1}(d), as compared to the quasi-dipole of Fig.~\ref{fig1}(f), is one example.
 Another example is that, in Fig.~\ref{fig1}(c), the strongest field does not arise around the 
 dominating current $\mathbf{J}_\theta$ ($R=d$), but for $R<d$ (notice the maxima in the figure take place for 
 lateral displacements smaller than $1$ nm from the center).
 The latter fact is related to the inverse proportionality between $|\mathbf{B}|$ and the loop radius $R$,
 which makes the inner loops have a significant contribution even if they host moderate current $\mathbf{J}_\theta$.

 \subsection{Quantum well}
\label{s:QW}

 A semiconductor quantum well (QW) embedded in a TI material involves two parallel planar interfaces.\cite{Harrison_book}
 As such, one can design the properties of the induced magnetic field from the superimposed effects
 of two simple planes, each with a different sign of $\boldsymbol{\nabla} \theta$.

\begin{figure}[htb]
\includegraphics[width=8cm]{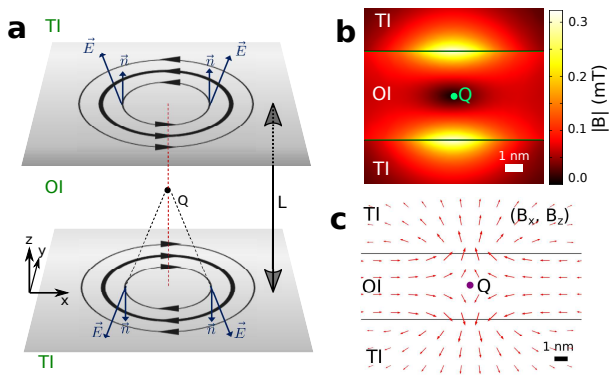}
\caption{TME in a QW with a centered charge.
(a) Schematic of the structure and the induced Hall currents $\mathbf{J}_\theta$.
Notice the opposite sense of rotation on top and bottom interfaces.
(b-c) numerical calculations for a QW with thickness $L=5$ nm.
%
%
(b) Absolute value of $\mathbf{B}$ on the $xz$ plane. The circle shows the position of the charge impurity.
(c) Logarithmic plot of the $(B_x,B_z)$ on the $xz$ plane. 
}
\label{fig2}
\end{figure}

 Consider first the high-symmetry case where the charge $Q$ is in the center of the QW, Figure \ref{fig2}(a).
 The Hall currents $\mathbf{J}_\theta$ on top and bottom interfaces have the same magnitude but opposite sense of rotation,
 set by the product $\mathbf{n} \times \mathbf{E}$.
 The result is that magnetic fields of identical magnitude but opposite sign are induced on each surface,
 and these can be expected to cancel out at the position of $Q$. This is indeed observed in actual calculations:
  $|\mathbf{B}|$ is suppressed around $Q$ --Figure \ref{fig2}(b)--,
 and the field distribution $(B_x,B_z)$ exhibits a nodal plane for $B_z$ at the center of the QW --Figure \ref{fig2}(c)--,
 which is reminiscent of that arising from two quasi-dipoles of opposite sign. 
 All of these are manifestations of the magnetoelectric interaction between the two QW planes,
 and set a first example on how nanostructures can be composed to design the resulting magnetic field.
 %

\begin{figure}[htb]
\includegraphics[width=8.0cm]{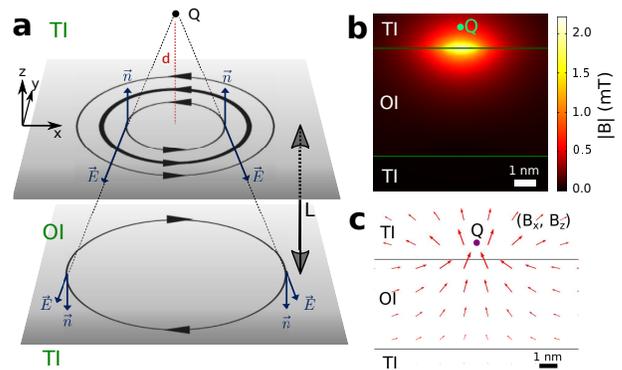}
\caption{TME in a QW with an off-centered charge.
(a) Schematic of the structure and the induced Hall currents $\mathbf{J}_\theta$.
The currents of the bottom interface provides a weak compensation to the top ones.
(b-c) numerical calculations for a QW with thickness $L=5$ nm and $d=1$ nm.
%
%
(b) Absolute value of $\mathbf{B}$ on the $xz$ plane. The circle shows the position of the charge impurity.
(c) Logarithmic plot of the $(B_x,B_z)$ on the $xz$ plane. 
}
\label{fig3}
\end{figure}

 We next consider the case where an off-centered charge $Q$ is placed in the TI material, 
 at a distance $d=1$ nm from the top of the QW.
 Figure \ref{fig3} analyzes the resulting magnetic field. The field is very similar to that
 of a single plane studied in Fig.~\ref{fig1}, because the electric field $\mathbf{E}$
 reaching the bottom interface is already weak, which reduces its influence. 
 A few signatures of the inter-plane interaction can however be observed. 
 Notice in Fig.~\ref{fig3}(c) that the field is severely quenched under the QW. 
 This is a consequences of the Hall currents on the bottom plane compensating for those
 on the top one.\cite{YngAPX,MoviXXX}

\begin{figure}[htb]
\includegraphics[width=8.0cm]{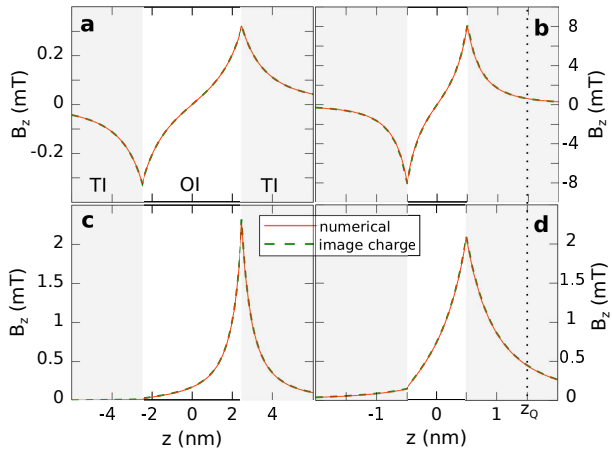}
\caption{
$B_z$ projection of the magnetic field along the $z$ axis for different QW thicknesses, $L$.
(a) and (b) QW with $Q$ centered at $z=0$, having $L=5$ nm and $L=1$ nm, respectively.
(c) and (d) same but with $Q$ on top of the QW, with $d=1$ nm.
Solid (dashed) lines are results calculated with numerical (image charges) method.
}
\label{fig4}
\end{figure}

 The results in Fig.~\ref{fig2} and \ref{fig3} correspond to a QW with thickness $L=5$ nm, 
which is a representative value for epitaxial wells\cite{DurnevPSS}. 
Qualitatively similar results are obtained for narrower QWs, 
such as colloidal nanoplatelets, which can be as thin as $L=1$ nm.\cite{DabardCM} 
 To illustrate this point, in Figure \ref{fig4} we compare the magnetic field $B_z$ along the $z$-axis
 for QWs of different $L$. 
 Fig.~\ref{fig4}(a) and (b) compare $B_z$ induced by a centered charge
 in QWs with $L=5$ nm and $L=1$ nm, respectively. In both cases, the field is antisymmetric
 with respect to the $z=0$ plane. The fields are stronger in the narrow QW  
  for the simple reason that $Q$ is closer to the interface.
  In the case of off-centered charges, Fig.~\ref{fig4}(c) and \ref{fig4}(d),
  since $d=1$ nm for both QW thicknesses, the maximum value of $B_z$ --around 2 mT--, 
  is fairly unsensitive to $L$.

 The accuracy of the numerical results in this section is again supported by the excellent
 agreement with analytical calculations using series of image charges\cite{YngAPX,MoviXXX}. 
 We compare the results of the two methods in Fig.~\ref{fig4}, using solid and dashed lines, 
 respectively.
 This provides further support to the reliability of the numerical integration for the more 
 elaborate nanostructures we address in the next sections.

\subsection{Quantum wire}
\label{s:QWW}

We consider now a cylindrical, semiconducting wire\cite{Harrison_book} with radius $R=2.5$ nm, embedded in a TI material.
As we shall see below, the curvature of the quantum wire (QWW) brings about some characteristic 
features, when compared to the QW.

\begin{figure}[htb]
\includegraphics[width=7.0cm]{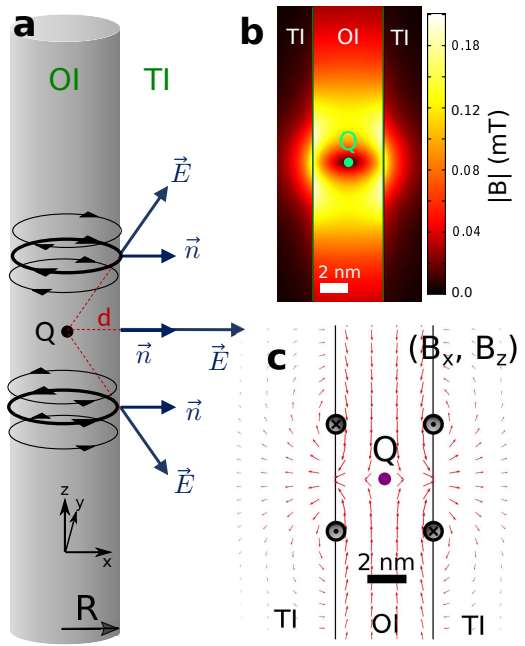}
\caption{ TME in a cylindrical QWW with a centered charge.
(a) Schematic of the structure and the induced Hall currents $\mathbf{J}_\theta$.
The currents above and below $Q$ have opposite senses of rotation.
(b-c) numerical calculations for a QWW with radius $R=2.5$ nm.
%
%
(b) Absolute value of $\mathbf{B}$ on the $xz$ plane. 
(c) Logarithmic plot of the $(B_x,B_z)$ on the $xz$ plane. 
The position of the strongest current loops is marked with $\odot$ and $\otimes$ symbols.
}
\label{fig5}
\end{figure}

In a QWW, the high symmetry configuration corresponds to the charge $Q$ placed at the center of the
circular cross section. As shown in Figure \ref{fig5}(a), no Hall currents are expected
at any point of the radial boundary of the $z=z_Q$ plane, because $\mathbf{n} \parallel \mathbf{E}$ 
and hence $\mathbf{J}_\theta=0$. By contrast, radial currents are formed at all distances above
and below $Q$. Interestingly, the product $\mathbf{n} \times \mathbf{E}$ leads to $\mathbf{J}_\theta$
having opposite sense of rotation in each case. This anticipates another suppression of $\mathbf{B}$
in the vicinity of $Q$, analogous to that of QWs, despite the different geometry.
This effect is confirmed by the numerical calculations.
In Figure \ref{fig5}(b), $|\mathbf{B}|$ shows a clear quenching around $Q$. 
At the same time, in Figure \ref{fig5}(c), the field orientation reveals that the $xz$ plane containing
$Q$ acts as a nodal plane for $B_z$.
 
The magnetic field generated by a centered charge in a QWW 
closely corresponds to that induced by two quasi-dipoles with opposite sign above and below $Q$.
 In Fig.~\ref{fig5}(b), two axially symmetric maxima of the field are observed at $z-z_Q \approx \pm R$. 
 At that position, in Fig.~\ref{fig5}(c), vortex-like circulations of the field 
 allow one to identify the positive and negative poles of the dipole, 
 which we represent using standard symbols of inbound ($\otimes$) and outbound ($\odot$) current.
It is inferred that $\mathbf{B}$ results mainly from two circular Hall current loops, which prevail over others.
The formation of dominating loops can be rationalized through the schematic in Figure \ref{fig5}(a).
As mentioned in Section \ref{s:plane}, the intensity of the currents $\mathbf{J}_\theta$ is set 
by the trade-off between the distance from $Q$, which weakens $|\mathbf{E}|$, 
and the angle between $\mathbf{n}$ and $\mathbf{E}$. As in planar interfaces,
this occurs for an angle of $\pm \pi/4$, which sets $z-z_Q \approx \pm R$. 
Unlike in planes and QWs, however, here all the current loops have the same radius $R$,
so the most intense currents translate into the strongest $|\mathbf{B}|$.

\begin{figure}[htb]
\includegraphics[width=7.0cm]{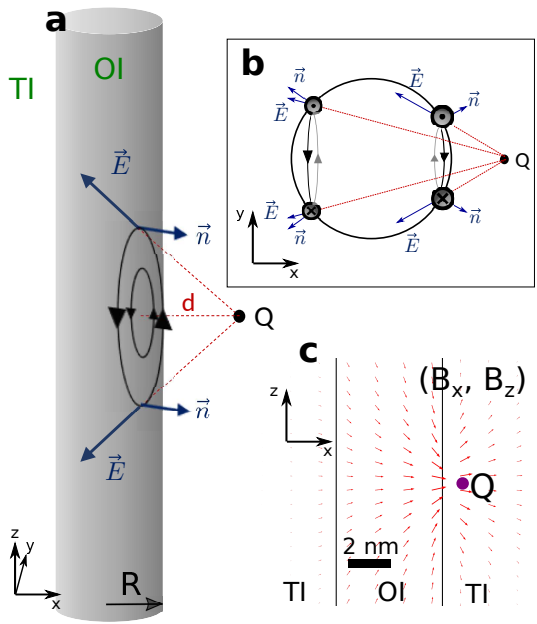}
\caption{TME in a QWW with an off-centered charge.
(a) Schematic of the structure and the induced Hall currents $\mathbf{J}_\theta$.
Non-circular currents loops are formed.
(b) Top view of the $xy$ plane containing $Q$.
(c) Logarithmic plot of the $(B_x,B_z)$ on the $xz$ plane, for a QWW with $R=2.5$ nm, and $d=1$ nm.}
\label{fig6}
\end{figure}

Figure \ref{fig6} analyzes the magnetic field induced by an off-centered charge $Q$, placed outside the QWW.
As shown in Fig.~\ref{fig6}(a), the Hall currents $\mathbf{J}_\theta$ are no longer circular. 
 Another characteristic feature is that, 
contrary to the planar interfaces of the QW, here the sign of the Hall currents on the front and back sides
of the QWW have the same sign. This can be understood from the diagram in Figure \ref{fig6}(b), which shows
a top view of the $xy$ cross-section containing $Q$. No matter the distance $d$ from $Q$ to the wire, the angle
between $\mathbf{n}$ and $\mathbf{E}$ has the same sign for front and back sides.
A direct consequence is that the magnetic field arising on the back interface does not compensate for that of the front one.
This implies that the calculated field, plotted in Fig.~\ref{fig6}(c), is not suppressed as the back interface is approached, 
and it can be felt on the opposite side of $Q$. This is in sharp contrast with the QW, Figure \ref{fig3}(c).
We stress that this behavior results from the circular curvature of the QWW cross-section, but increasing the ellipticity
should eventually retrieve the QW behavior. Therefore, one can anticipate the existence of a critical eccentricity at
which $\mathbf{E} \parallel \mathbf{n}$ on the back interface, at which the $\mathbf{B}$ sign is reversed.
%
%
 A related mechanism has been put forward for semi-spherical cavities, 
where the sign of the magnetic field can be reversed through the distance $d$ between $Q$ and the center of the sphere.\cite{CamposPLA}

\subsection{Quantum dot}
\label{s:QD}

Quantum dots (QDs) exhibit nanometric confinement in all three directions of space.\cite{Harrison_book}
When embedded inside a TI, this involves magnetoelectrically polarized boundaries, 
all within short distances from each other. Sizable interactions may then take build up within them. 
Combined with the precise synthetic control of their size and shape,\cite{KovalenkoACS} 
these properties make QD systems particularly suited to modulate the TME.
 To investigate such interactions, here we consider two basic geometries, namely spherical and cuboidal QDs.
 QDs with geometries resembling these limit cases are now routinely achieved with colloidal chemistry.\cite{KovalenkoACS}

\begin{figure}[htb]
\includegraphics[width=8.0cm]{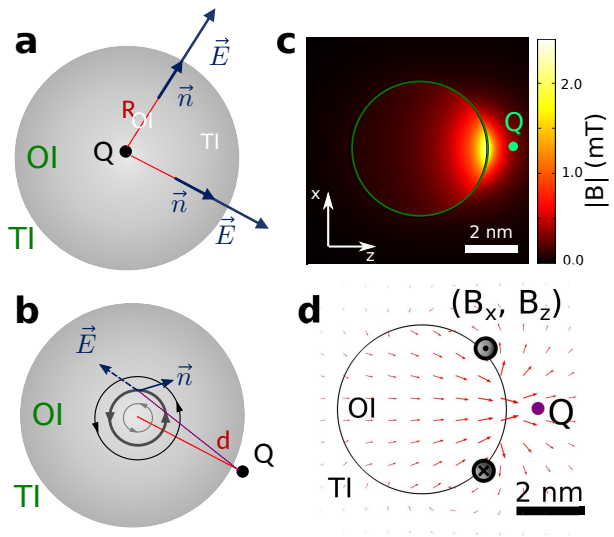}
\caption{TME in a spherical QD. 
(a) Schematic of a QD with a centered charge. No Hall currents are formed, and hence no magnetic field is induced.
(b) Same but for an off-centered charge, within the TI.
(c-d) Numerical calculations for a QD with $R=2.5$ nm and $d=1$ nm.
(c) Absolute value of $\mathbf{B}$ on the $xz$ plane. 
(d) Logarithmic plot of the $(B_x,B_z)$ on the $xz$ plane, induced by the off-centered charge.}
\label{fig7}
\end{figure}

 Figure \ref{fig7} shows results for a spherical QD. If the source charge $Q$ is centered, 
 the spherical symmetry of the system leads to $\mathbf{n} \parallel \mathbf{E}$ all over the boundary, see Fig.~\ref{fig7}(a).
 Therefore, no Hall currents are formed, and no magnetic field is induced. 
 The suppression of the TME is however lifted as soon as the charge is off-centered.
 Figure \ref{fig7}(b) shows the expected Hall currents triggered by a charge outside the QD.
 These are spherical and, as in the QWW, have the same sign on the front and back sides of the nanostructure.
 The ensuing magnetic field intensity is strongest near the QD cap close to the charge, Fig.~\ref{fig7}(c),
 and the field distribution, plotted in Fig.~\ref{fig7}(d), exhibits circulating vortices, 
 which makes it reminiscent of a simple quasi-dipole field. 
 This behavior is in contrast to that of a planar interface, Fig.~\ref{fig1}(d), 
 and reflects the fact that interferences between concentric Hall currents on the sphere surface are less
 destructive than those on a plane.

\begin{figure}[htb]
\includegraphics[width=12.0cm]{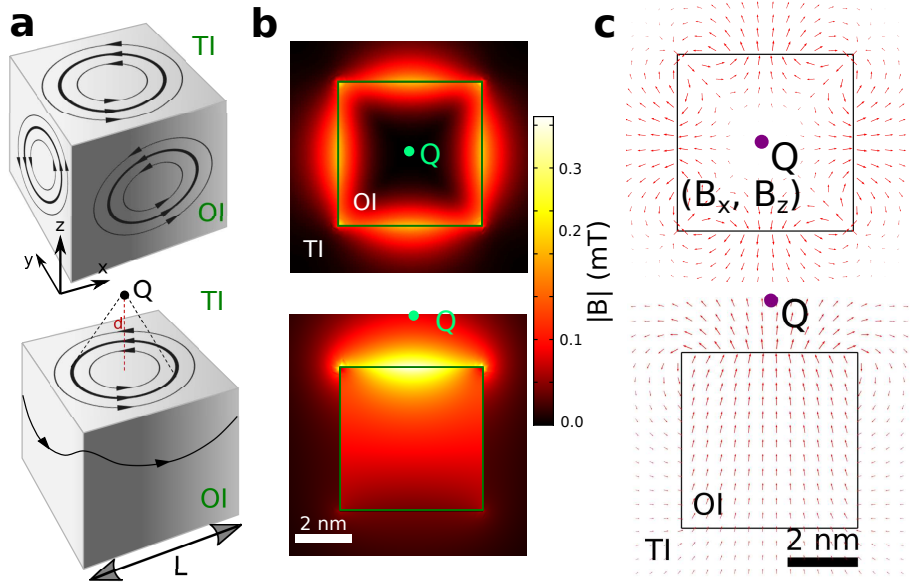}
\caption{TME in a cuboidal QD with centered (top row) and off-centered (bottom row) charge.
(a) Schematic of the structure and the resulting Hall currents. 
(b-c) Numerical calculations for a QD with $L=5$ nm and $d=2.5$ nm (off-centered charge).
(b) $|\mathbf{B}|$ on the cube plane containing the charge $Q$.
(c) Corresponding plots of field distribution $(B_x, B_z)$.}
\label{fig8}
\end{figure}

Cuboidal QDs are a paradigmatic example on how the proximity of boundaries in QDs can be manipulated to shape the TME.
Compare the Hall currents $\mathbf{J}_\theta$ induced by a charge $Q$ in the center of a cube or outside it,
top and bottom rows of Figure \ref{fig8}(a), respectively. 
 In the former case (top schematic), circular currents with opposite sign form on each pair of  
parallel faces. These tend to compensate each other, similar to the QW case --Fig.~\ref{fig2}(a)--. 
%
 On the contrary, if the charge is outside (bottom schematic), the Hall currents are similar to those of a simple plane, 
 but with extra currents with same sense of rotation formed on the lateral sides, which reinforce the resulting
 magnetic field.
 The constrasting behavior is reflected in Figure \ref{fig8}(b), where $|\mathbf{B}|$ is plotted for a 
 cross-section of the QD containing the charge $Q$. The cube side is $L=5$ nm and, for the sake of comparison,
 $Q$ is placed at $d=2.5$ nm from the surface both in the centered and off-centered configurations.
 When $Q$ is centered (top panel), a broad area with quenched magnetic field is formed. 
 On the contrary, when $Q$ is off-centered (bottom panel), the resulting field becomes stronger, 
 and a slow decay is observed inside the QD, which is in contrast with the fast decay outside the QD. 
 Both the quenching and the enhancement of the magnetic field are 
 more efficient than those observed in QWs of similar dimensions (cf.~Fig.~\ref{fig2}(b) and \ref{fig3}(b)). 
 Drastic changes are seen in the field distribution as well depending on the charge location, 
 owing to the cuboidal geometry, see Figure \ref{fig8}(c), with a prominent role played by the cube edges.

 \onecolumngrid

\twocolumngrid

\subsection{Quantum ring}
\label{s:QR}

Doubly connected, ring-like semiconductor nanostructures have attracted attention in the solid state community
for their ability to reveal topological phenomena, such as the Aharonov-Bohm effect in the presence of external
magnetic fields.\cite{FuhrerNAT,Fomin_book}
 We consider here an ideal, torus-shaped quantum ring (QR). 
 Because the QR can be seen as a bended wire, the magnetic fields induced by off-centered charges close to one side 
 of the QR are similar to those of QWW, albeit with deviations arising from the curvature of the ring arm (not shown).
 If the charge $Q$ is centered, however, the analogy is less straightforward.
 
\begin{figure}[htb]
\includegraphics[width=6.5cm]{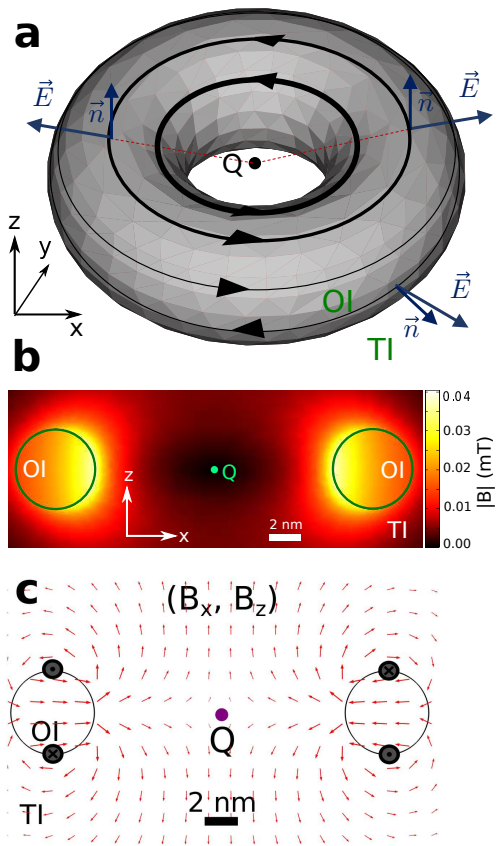}
\caption{ TME in a QR with a centered charge.
(a) Schematic of the structure and the induced Hall currents $\mathbf{J}_\theta$.
The currents above and below $Q$ have opposite senses of rotation and circulate around $Q$.
(b-c) numerical calculations for a toroidal QR with major radius 10 nm and minor (section) radius $2.5$ nm.
%
%
(b) Absolute value of $\mathbf{B}$ on the $xz$ plane. 
(c) Logarithmic plot of the $(B_x,B_z)$ on the $xz$ plane. 
The position of the strongest current loops is marked with $\odot$ and $\otimes$ symbols.
}
\label{fig9}
\end{figure}

 Analyzing the Hall currents $\mathbf{J}_\theta$ induced by a centered charge, Figure \ref{fig9}(a), one infers
 that these spin in circles around $Q$, with opposite sense of rotation above and below the $z_Q$ (equatorial plane),
 which is again reminiscent of the cylindrical QWW. 
 %
 This behavior is confirmed by the calculated field distribution, plotted in Figure \ref{fig9}(c), which corresponds 
 roughly to that set by two dominant quasi-dipoles, marked by $\odot$ and $\otimes$ symbols in the figure.
 The field intensity, however, is strongest on the inner side of the QR, Figure \ref{fig9}(b).
 In this regard, the behavior is closer to that of an off-centered charge near a spherical QD, Fig.~\ref{fig7}(b).

\section{Conclusions}

We have introduced a simple numerical model to calculate magnetoelectric fields generated by electrostatic charges in TI-OI heterostructures of arbitrary geometry.
The model is quantitatively as accurate as the image charge method, but with the advantadge of providing easy access to non-planar and elaborate geometries. It thus constitutes a fast and versatile tool to build models pursuing specific effects.
 We have then used the model to invesigate the TME in the most common semiconductor nanostructures: quantum dots, rings, wells and wires.
By changing the position of a source electric charge with respect to the nanostructure, the strength and orientation of the resulting magnetic field undergoes severe changes.
These results confirm that low-dimensional nanostructures, embedded in TI media, constitute a particularly rich playground to investigate and exploit the TME.

The most conspicuous influence of the nanostructure geometry is observed in cuboidal QDs, where the interactions between boundaries can be used to efficiently
enhance or quench the induced magnetic fields.

.

\begin{acknowledgments}
We acknowledge support from MICINN project PID2021-128659NB-I00, 
UJI project B-2021-06 and Generalitat Valenciana Prometeo project 22I235-CIPROM/2021/078.
\end{acknowledgments}

\bibliographystyle{apsrev4-1}
\bibliography{topobib}


\end{document}